\documentclass[a4paper,11pt]{article}
\pdfoutput=1 
\usepackage{jinstpub}
\usepackage[utf8]{inputenc}

\title{Visualization of radiotracers for SPECT imaging using a Timepix detector with a coded aperture}

\author[a]{V.~Rozhkov}
\author[a]{, G.~Chelkov}
\author[b]{, I.~Hern\'{a}ndez}
\author[c]{, O.~Ivanov}
\author[a]{, D. Kozhevnikov}
\author[a,d]{, A.~Leyva}
\author[b]{, A.~Perera}
\author[a,e]{, D.~Rastorguev}
\author[a]{, P.~Smolyanskiy}
\author[b]{, L.~Torres}
\author[a]{, A.~Zhemchugov}

\affiliation[a]{Joint Institute for Nuclear Research, Dubna, Russia}
\affiliation[b]{Isotopes Center, Mayabeque, Cuba}
\affiliation[c]{National Research Center «Kurchatov Institute», Moscow, Russia}
\affiliation[d]{Center of Technological Applications and Nuclear Development, Havana, Cuba}
\affiliation[e]{Moscow Institute of Physics and Technology, Dolgoprudny, Russia}

\emailAdd{rozhkov@jinr.ru}

\abstract{The work shows the ability to visualize radiotracers used in SPECT with a system based on a coded aperture mask and a hybrid pixel Timepix detector with the CdTe sensor. Characterization of the system using X-rays and radioactive sources confirms that the spatial resolution of less than 1~mm with a field of view 3~cm $\times$ 3~cm can be achieved. The results of a simulation study to determine the expected spatial resolution of the system in the focal plane for the various radionuclides is presented. The possibility of using this system with a thin (1.5~mm) coded aperture mask for reconstructing images of gamma emitters with the energy up to 180~keV is demonstrated.}


\begin{document}
\maketitle
\flushbottom

\def\figurename{Fig.}
\def\tablename{Table}

\section{Introduction}
\label{sec:intro}
\vspace{\baselineskip}
\par The visualization of internal organs of small animals {\it in vivo} has become one of the main tasks in preclinical studies over the last decade~\cite{1,2,3}. Single-photon emission computed tomography (SPECT) allows obtaining tomographic images of the biodistribution of radiolabeled compounds, both throughout the patient’s body and in separate organs. Radiopharmaceuticals labeled with some gamma-emitter radionuclides (E$\gamma$=80-350~keV) are used in SPECT, in contrast to the positron emission tomography (PET). SPECT is currently one of the most effective and highly sensitive imaging methods to study the function of internal organs and tissues, as well as a key tool in the development of new radiopharmaceuticals and to seek for methods for their targeted delivery~\cite{4,5}. Small animal imaging techniques also help to reduce the number of animals in non-clinical research providing a way to observe radiopharmaceuticals and other drugs {\it in vivo} distribution in a noninvasive manner.

\par In the traditional gamma-ray registration device (Anger gamma camera), the detecting unit consists of a collimator, scintillator and photomultiplier tubes (PMTs). Gamma radiation passing through the collimator interacts with the scintillation crystal, where the emitted energy is converted into the visible light, subsequently detected by the PMT. The main disadvantage of the gamma camera is its relatively low spatial resolution, which typically equals to several millimeters or more and which is limited by the collimator construction~\cite{6}. 

\par In studies of small animals, the region of interest typically has a small size and a high spatial resolution is necessary to get a good image. However, the increase of the spatial resolution with the given specific activity of radiopharmaceutical results in larger statistical fluctuations in the image, especially if the detection efficiency is not high enough. This could be compensated by the increase of the activity of the radiotracer and the exposure time, but only up to a certain limit determined by the maximum allowed dose and the mobility of the studied object (breathing, occasional movements etc)~\cite{5}. This makes good detection efficiency to be another key property of the SPECT system.

\par With the exception to the Compton camera~\cite{7,8}, the main way to get an image in SPECT is to collimate the gamma-rays. The quality of the image largely depends on the choice of collimator. The spatial resolution of a gamma camera can be improved using a pin-hole and a high-resolution coordinate detector~\cite{9}. However, this method is not efficient since only those photons that pass through the hole will be detected and the rest of them will be absorbed by the collimator. The larger is the aperture of the pin-hole, the higher is the sensitivity. However, the spatial resolution of the system deteriorates with the increase of the pin-hole aperture~\cite{10}.

 \par An alternative is to use a coded aperture (CA) mask~\cite{11}. It is a perforated plate with a number of holes (transparent elements), where the location of holes follows the specific pattern. The plate is made of a heavy material to reach a high gamma absorption. Opposite to a pin-hole collimator, the CA combines the high spatial resolution with the high sensitivity. In the case of CA, the image of an object in the detector plane will be a result of the superposition of the images formed by each hole (a shadowgram). The image of the real object can be reconstructed from the shadowgram of $N \times N$ dimension using convolution with a decoding function~\cite{12}:
\begin{equation}
\label{eq:1}
\begin{split} I_{\rm k, l} = \sum_{\rm j=1}^{\rm N}\sum_{\rm i=1}^{\rm N}D_{\rm i, j}\cdot M_{\rm i+k, j+l},\end{split}
\end{equation}
where $I_{\rm k,l}$ is an element of the reconstructed image, $D_{\rm i, j}$ is an element of the shadowgram and $M$ is the decoding function.
\par 
\section{Experimental setup}
\vspace{\baselineskip}

\par The goal of this work is to study the performance with a  system based on a coded aperture mask and a hybrid pixel Timepix detector with the CdTe sensor~\cite{13,14,15} in SPECT applications, aimed at the imaging of small animals. A series of experiments with various radiation sources was carried out with the following setup~(Fig.\ref{fig:1}):

\begin{figure}[!ht]
\centering 

\qquad
\includegraphics[scale = 0.5]{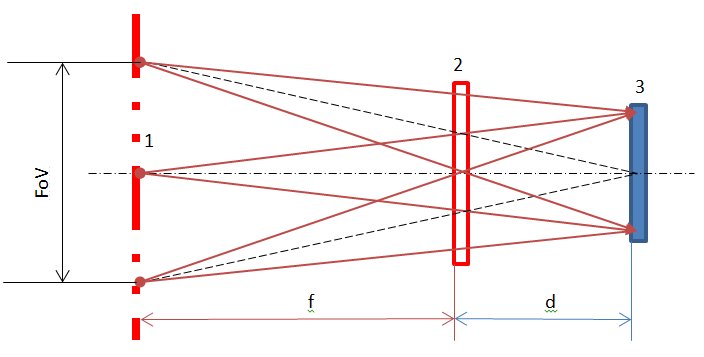}
\caption {\label {fig:1} Layout of the experimental setup:
1~–~source plane, 2 – coded aperture, 3 – detector, {\it FoV} – field of view,{\it f}~–~distance from the source to the coded aperture, {\it d} – distance from the coded aperture to the detector}

\end{figure}

\begin{itemize}
\item \textbf{Detector}: a hybrid pixel detector based on the Timepix readout chip  with a 1~mm thick CdTe sensor. The main properties of the detector are summarized in Table~\ref{tab:1}. The detector is capable to record the position of a gamma ray interaction and to determine the energy deposit in the sensor for every particle~\cite{16}. Each particle interaction may induce a signal in one or several adjacent detector pixels, thus forming a cluster. Weighted average of all signals in the cluster provides the coordinate of the particle interaction and the total amplitude of all signals in the cluster is proportional to the energy deposit in the sensor. During the exposure, clusters are recorded and then only those are used further whose energy falls in the certain range. Position of the selected clusters is used to form an 256 pixels wide by 256 pixels high image, where every cluster enters with the unit weight.

\begin{table}[!ht]
\centering
\caption{\label{tab:1} Detector properties.}
\smallskip
\begin{tabular}{|lr|c|}
\hline
Sensor material &	CdTe \\
\hline
Sensor size	& 14.1~mm~x~14.1~mm \\
\hline
Sensor thickness	& 1~mm \\
\hline
Pixel matrix &	256~x~256 \\
\hline
Pixel size &	55~$\rm \mu$m~x~55~$\rm \mu$m \\ 
\hline
Energy resolution for 60~keV gamma rays \cite{17} &	5.6\% \\ 
\hline
Gamma ray detection efficiency\cite{18} &	$\approx$100\% below 60~keV, 65\% at 100~keV\\
\hline
\end{tabular}
\end{table}

\item \textbf{Collimator}: a set of identical tungsten CA masks with holes following the square MURA pattern with the rank equal to 31. The dimension of workarea is 22~mm~$\times$~22~mm. The radius of a hole is of 170$\pm$10~$\rm \mu$m. The thickness of every mask is equal to 0.5~mm. Depending on the energy of gamma rays, a set can consist of more than one mask to improve the absorption. However, a thick mask decreases the {\it FoV} and deteriorates the reconstruction near the edges. Therefore, the optimal thickness is a trade-off between the collimator absorption and the spatial resolution of the measuring system, which is especially important for the multiple hole collimator with a small diameter of a hole~\cite{19,20}.

\begin{figure}[!ht]
\centering 
\includegraphics[scale = 0.05]{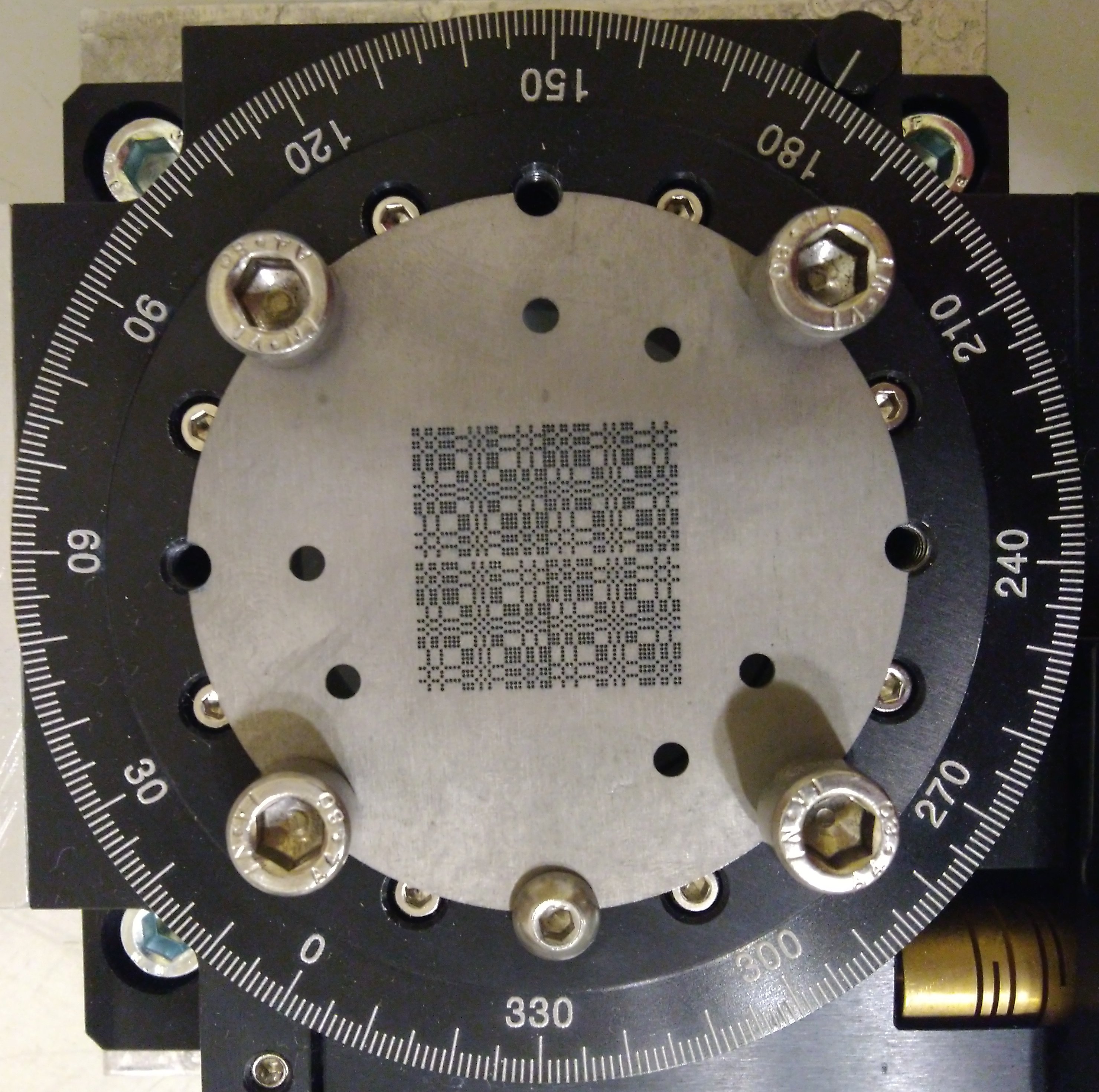}
\centering 
\caption{\label{fig:2} Mounting of the CA to a turntable.}
\end{figure}

\par In this study a set of three CA masks was used with a total thickness of 1.5~mm (which is enough to attenuate the intensity of 160~keV gamma rays by a factor of 30) fixed on a vertically mounted turntable, thereby ensuring the rotation of the masks around their center~(Fig.\ref{fig:2}). A distinctive feature of MURA type CA is that the rotation of a mask by 90$^\circ$ closes previously open mask elements and open the closed ones. This allows to reduce the systematic background, thereby increasing the signal-to-noise ratio~\cite{11,12}.

\item \textbf{Sources}: three radiation sources have been used for the detector characterization. Namely, a microfocus X-ray source\footnote{X-ray source SB~120-350 by SourceRay Inc.}, the 59.5~keV gamma rays from the $^{241}Am$ radioactive source with the activity of 0.1~MBq and the diameter of approximately~2 mm, and various phantoms filled with~$^{99m}Tc$. \end{itemize}


\par
\begin{figure}[!ht]
\centering 
\includegraphics[scale = 0.45]{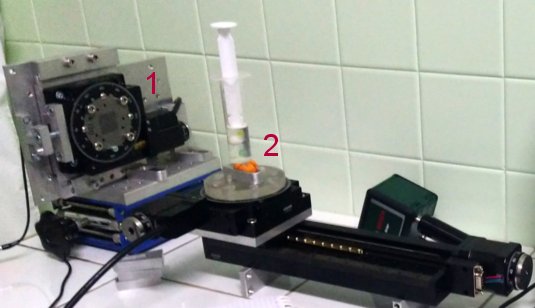}
\caption{\label{fig:3} A general view of the experimental setup. 1 - The Timepix detector with the CA mask installed, 2 - an object under study.}
\end{figure}

 The values of {\it f} and {\it d} were chosen so that the field of view was about 3~cm~$\times$~3~cm and the shadowgram from the base mask pattern completely fitted to the detector area, which was the necessary condition for the unambiguous image reconstruction from the shadowgram. A general view of the experimental setup is shown in~Fig.\ref{fig:3}.

\section{Measurement of the spatial resolution}
\vspace{\baselineskip}
\par The system spatial resolution of the detector was measured with the CA mask installed. Two types of sources were used. First, a point-like X-ray source and a thin linear source were used to directly measure the FWHM of the point spread function (PSF) and the line spread function (LSF), respectively. Second, small radioactive objects comparable in size with the expected spatial resolution were used as a cross-check. In the latter case, the response function was fitted by a convolution of the uniform distribution and a Gaussian and the spatial resolution was obtained using the standard deviation $\sigma$~\cite{21}:
\begin{equation}
\label{eq:2}
\begin{split} FWHM = \sigma * 2.35 \end{split}
\end{equation}
\par
\begin{figure}[!ht]
\begin{minipage}[h]{0.45\linewidth}
\center{\includegraphics[scale=0.5]{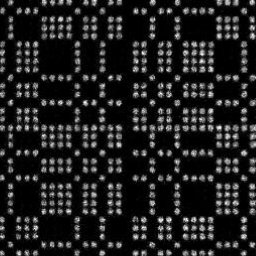}} \\a) \\
\end{minipage}
\hfill
\begin{minipage}[h]{0.45\linewidth}
\center{\includegraphics[scale=0.5]{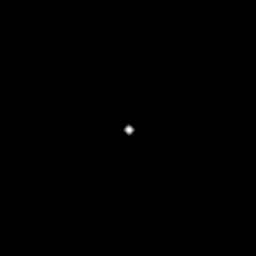}} \\b)
\end{minipage}
\vfill
\begin{minipage}[h]{1\linewidth}
\center{\includegraphics[scale=0.25]{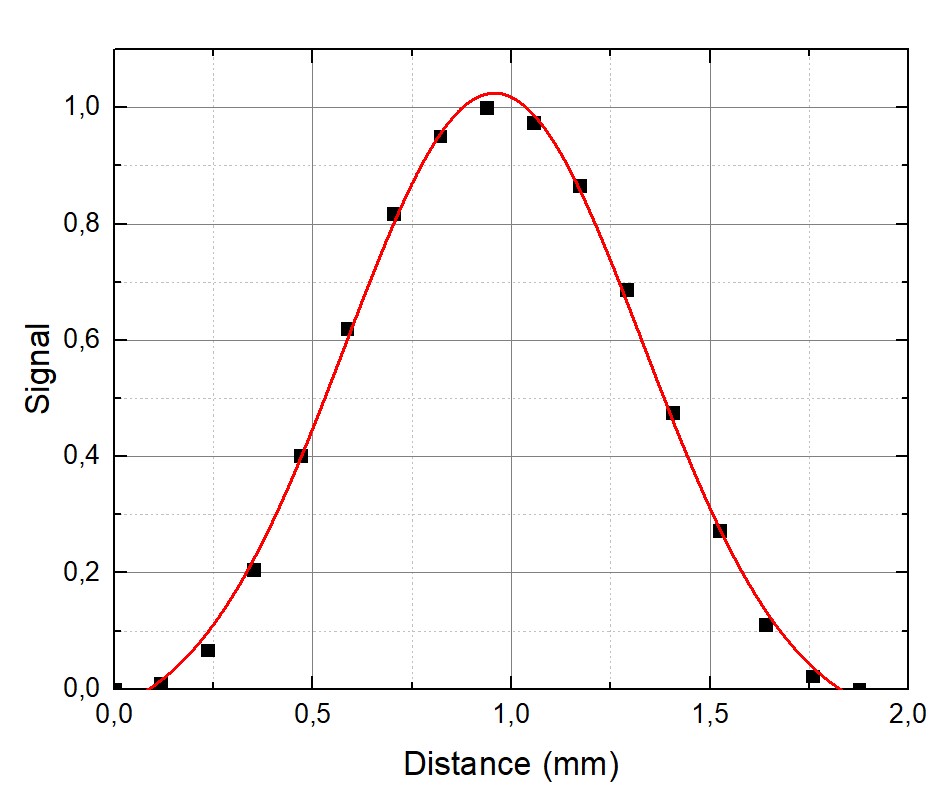}} \\c) \\
\end{minipage}
\caption{The point-like source (selected energy range: 6 keV - 60 keV):
a) the shadowgram obtained using the X-ray source, b) the reconstructed image, c) the response function to the point-like source.}
\label{fig:4}
\end{figure}

 The X-ray source with voltage of 60~kV and the tube current of 100~$\rm \mu$A was used to measure the PSF. The focal spot size of 175~$\rm \mu$m could be considered point-like. The photon energy followed the continuous spectrum with maximum at 30 keV. The shadowgram obtained with the 1~s exposure, the reconstructed image and the response function to the point-like source are shown in Fig.~\ref{fig:4}. The FWHM of the response function equals to 0.88~mm.

\par
\begin{figure}[!ht]
\begin{minipage}[h]{0.35\linewidth}
\center{\includegraphics[scale=0.45]{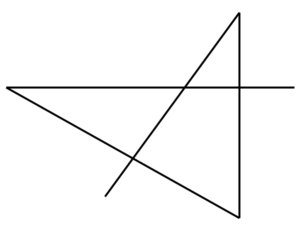}} \\a) \\
\end{minipage}
\begin{minipage}[h]{0.25\linewidth}
\center{\includegraphics[scale=0.5]{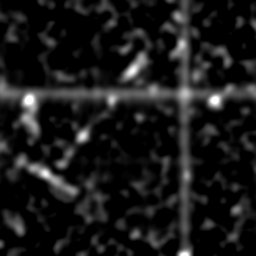}} \\b) \\
\end{minipage}
\begin{minipage}[h]{0.4\linewidth}
\center{\includegraphics[scale=0.2]{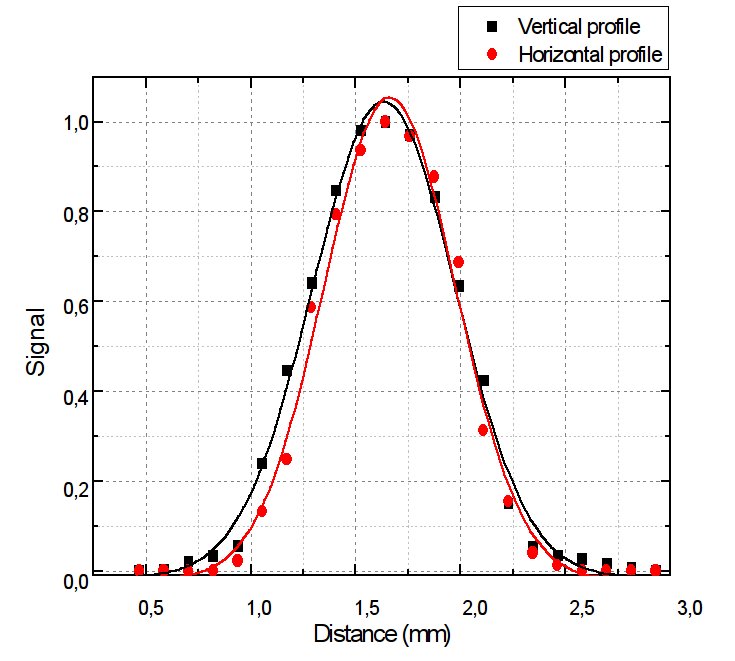}} \\c) \\
\end{minipage}
\caption{Visualization of a thread saturated with $^{99m}Tc$ (selected energy range: 90~keV - 145~keV): a) a thread pattern, b) the reconstructed image, c) vertical and horizontal LTFs.}
\label{fig:5}
\end{figure}

 A 150~$\rm \mu$m thick cotton thread saturated with a solution containing $^{99m}Tc$ was used to determine the LSF. The thread was stretched in a pattern that formed intersections at 90$^\circ$ and about 45$^\circ$, as shown in Fig.~\ref{fig:5}a. The soaking and drying caused an uneven distribution of $^{99m}Tc$ along the thread (Fig.~\ref{fig:5}b). However, the LSF of reconstructed horizontal and vertical thread images was measured with the FWHM to be 0.75~mm and 0.80~mm, respectively (see Fig.~\ref{fig:5}c). 

\par
\begin{figure}[!ht]
\begin{minipage}[h]{0.3\linewidth}
\center{\includegraphics[scale=0.9]{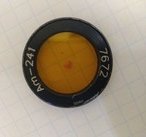}} \\a) \\
\end{minipage}
\begin{minipage}[h]{0.3\linewidth}
\center{\includegraphics[scale=0.47]{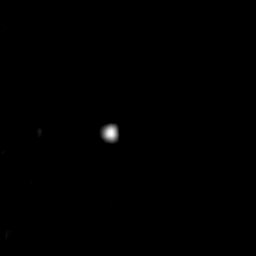}} \\b)
\end{minipage}
\begin{minipage}[h]{0.4\linewidth}
\center{\includegraphics[scale=0.28]{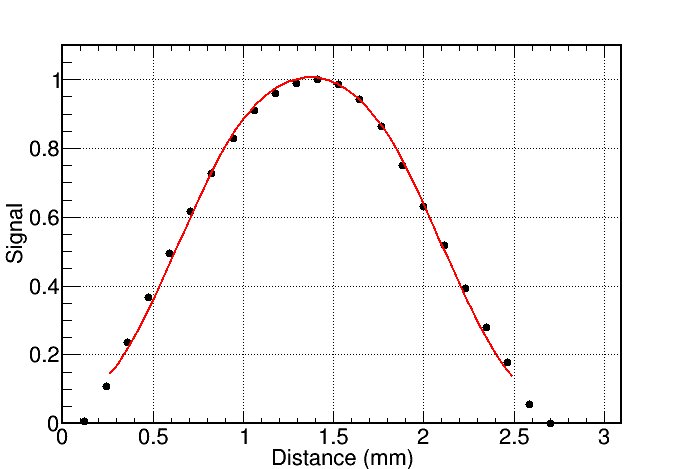}} \\c) \\
\end{minipage}
\caption{$^{241}Am$ spectrometric source (selected energy range: 40 keV - 60 keV): 
a) $^{241}Am$ spectrometric source, b) the reconstructed image, c) the response function to the $^{241}Am$ spectrometric source.}
\label{fig:6}
\end{figure}

 A spectrometric gamma-ray source of $^{241}Am$ of small size (Fig.~\ref{fig:6}a) was used as a cross-check. The reconstructed image is shown in Fig.~\ref{fig:6}b. The image profile was fitted by a convolution of the uniform distribution with the width of 1.46~mm and a Gaussian with standard deviation of 0.35~mm (see Fig.~\ref{fig:6}c). Therefore, the spatial resolution was equal to 0.82~mm (Fig.~\ref{fig:6}b,c), which was consistent with the results obtained using the point-like and the thin linear sources.

\par
\begin{figure}[!ht]
\begin{minipage}[h]{0.33\linewidth}
\center{\includegraphics[scale=0.03]{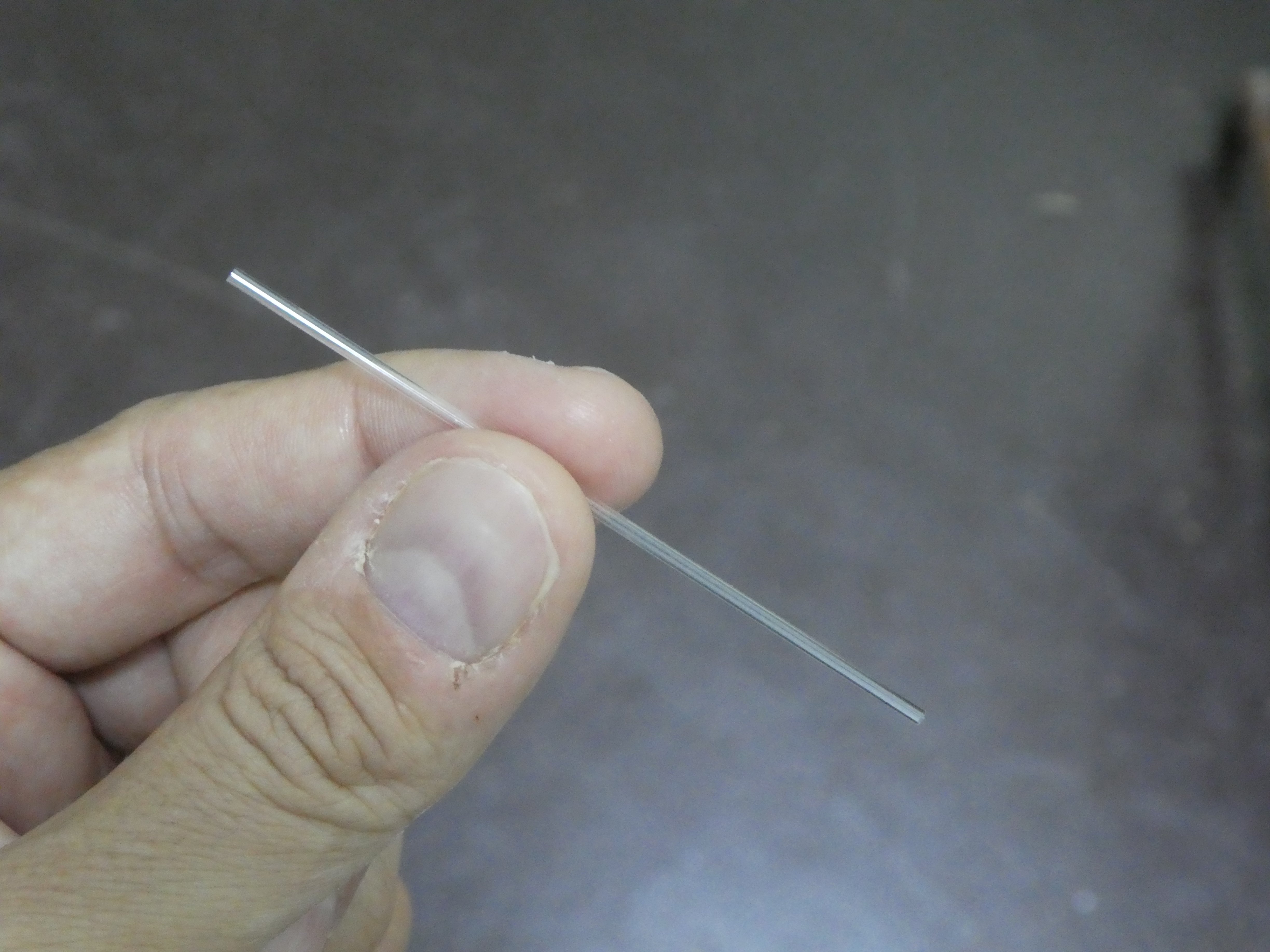}} \\a) \\
\end{minipage}
\begin{minipage}[h]{0.33\linewidth}
\center{\includegraphics[scale=0.46]{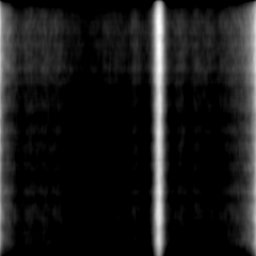}} \\b)
\end{minipage}
\begin{minipage}[h]{0.33\linewidth}
\center{\includegraphics[scale=0.46]{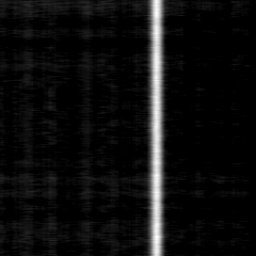}} \\c) \\
\end{minipage}
\caption{Capillary with $^{99m}Tc$ (selected energy range: 90~keV - 145~keV): 
a) the picture of the capillary, b) the reconstructed image, c) the  simulation.}
\label{fig:7}
\end{figure}

\begin{figure}[!ht]
\centering 
\qquad
\includegraphics[scale = 0.25]{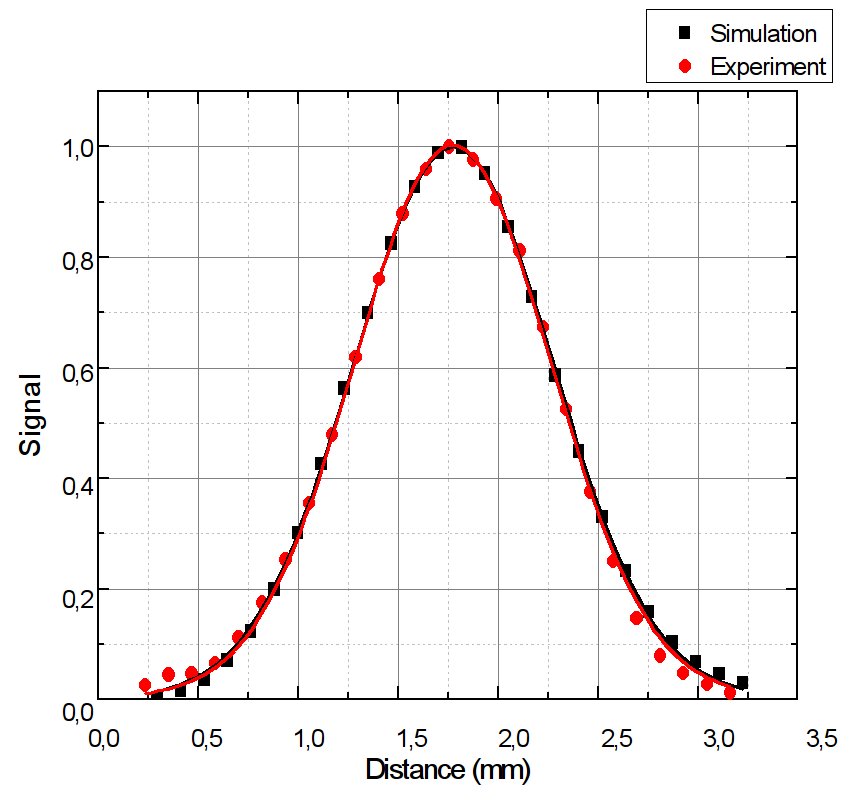}
\centering 
\caption{\label{fig:8} Comparison of experimental and simulated profiles of the capillary filled with $^{99m}Tc$.}
\end{figure}

 Another cross-check was made using a capillary filled with a solution containing $^{99m}Tc$. The internal and external diameter of the capillary was 1~mm and 1.5~mm, respectively. The $^{99m}Tc$ activity was 156~MBq and the exposure time was 2 minutes. The picture of the capillary and the reconstructed image are shown in Fig.~\ref{fig:7}a and Fig.~\ref{fig:7}b. The reconstructed image profile was fitted by a convolution of 1~mm wide rectangular distribution with a Gaussian. The spatial resolution of 0.74~mm was derived from the fit, that confirmed the results obtained using the thin linear source (see Fig.~\ref{fig:8}).

\par
\begin{figure}[!ht]
\begin{minipage}[h]{0.3\linewidth}
\center{\includegraphics[scale=1]{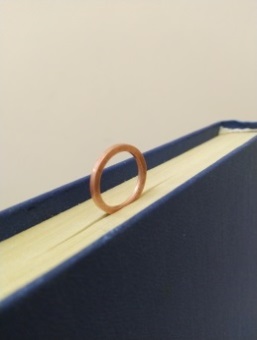}} \\a) \\
\end{minipage}
\hfill
\begin{minipage}[h]{0.34\linewidth}
\center{\includegraphics[scale=0.6]{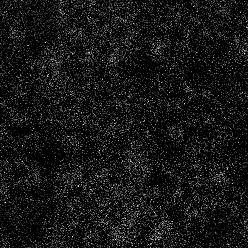}} \\b)
\end{minipage}
\hfill
\begin{minipage}[h]{0.3\linewidth}
\center{\includegraphics[scale=0.575]{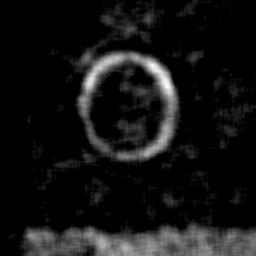}} \\c) \\
\end{minipage}
\vfill
\begin{minipage}[h]{1\linewidth}
\center{\includegraphics[scale=0.45]{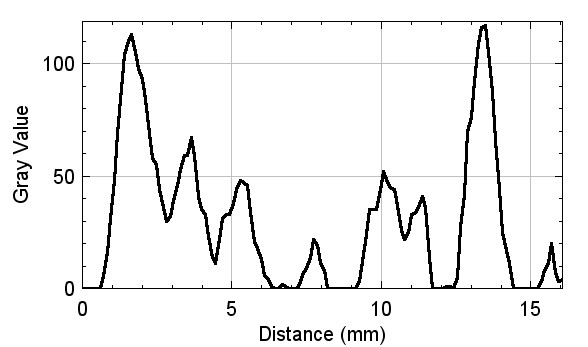}} \\d)
\end{minipage}
\caption{Visualization of the copper ring (selected energy range: 6~keV - 9~keV): a) a picture of the copper ring, b) the shadowgram, c) the reconstructed image, d) the profile.}
\label{fig:9}
\end{figure}

 Finally, the characteristic radiation emitted from a copper ring irradiated by X-rays was used to obtain the image of the complex shape (K$\alpha$ line of 8.98~keV was used). The thickness of the ring was 1~mm and the external and internal diameters -- 11.9~mm and 10.2~mm, respectively. The ring was rotated by 45$^\circ$ with respect to the detector plane. The results are shown in Fig.~\ref{fig:9}. The average diameter of the ring from the reconstructed image is equal to 11.4~mm, which is compatible with the real diameter of the ring.

\section{Simulation study}
\vspace{\baselineskip}
\par  The systematic study of the response of the system based on coded aperture to the gamma sources of different shape, with the energy up to 350~keV was carried out using the simulation based on the Geant4 toolkit~\cite{22,23}. The experimental setup with 1.5~mm thick CA mask was modelled. The simulation did not take into account the effects associated with the charge collection in the sensor and the readout electronics of the Timepix chip. Instead, the true position of the particle incident was taken into consideration and the simulated energy deposit was smeared with the experimental energy resolution.  Low energy Geant4 electromagnetic package~\cite{24}, based on the Livermore data libraries, was used to simulate photon interactions with the expected cross-section precision within 10\% in the energy range of interest~\cite{25}. The experimental data obtained with $^{241}Am$ were used to adjust the geometrical parameters of the model. Other experimental data sets were used to cross-check the simulation up to the photon energy of 140~keV. Prediction at higher energy relied on the validity of physics models of Geant4 corroborated by the cross-check results. The reconstruction algorithm was identical to the one used in the experiment, including the selection of the energy range.

\par
\begin{figure}[!ht]
\begin{minipage}[h]{0.5\linewidth}
\center{\includegraphics[scale=0.4]{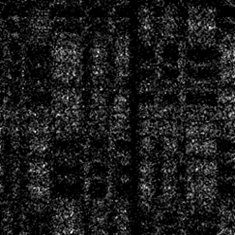}} \\
\end{minipage}
\hfill
\begin{minipage}[h]{0.5\linewidth}
\center{\includegraphics[scale=0.4]{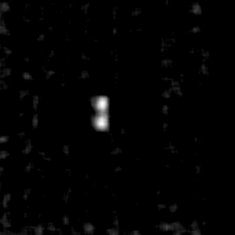}} \\
\end{minipage}
\vfill
\begin{minipage}[h]{0.5\linewidth}
\center{\includegraphics[scale=0.4]{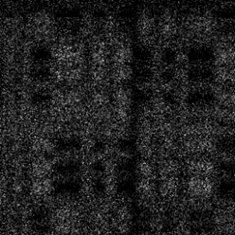}} \\
\end{minipage}
\hfill
\begin{minipage}[h]{0.5\linewidth}
\center{\includegraphics[scale=0.4]{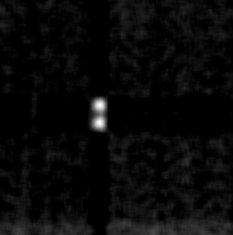}} \\
\end{minipage}
\caption{Visualization of two $^{241}Am$ sources (selected energy range: 50~keV - 65~keV): shadowograms (left) and reconstructed images (right) are shown. Simulated data are plotted in the top and the experimental data -- in the bottom.}
\label{fig:10}
\end{figure}

\begin{figure}[!ht]
\centering 
\qquad
\includegraphics[scale = 0.25]{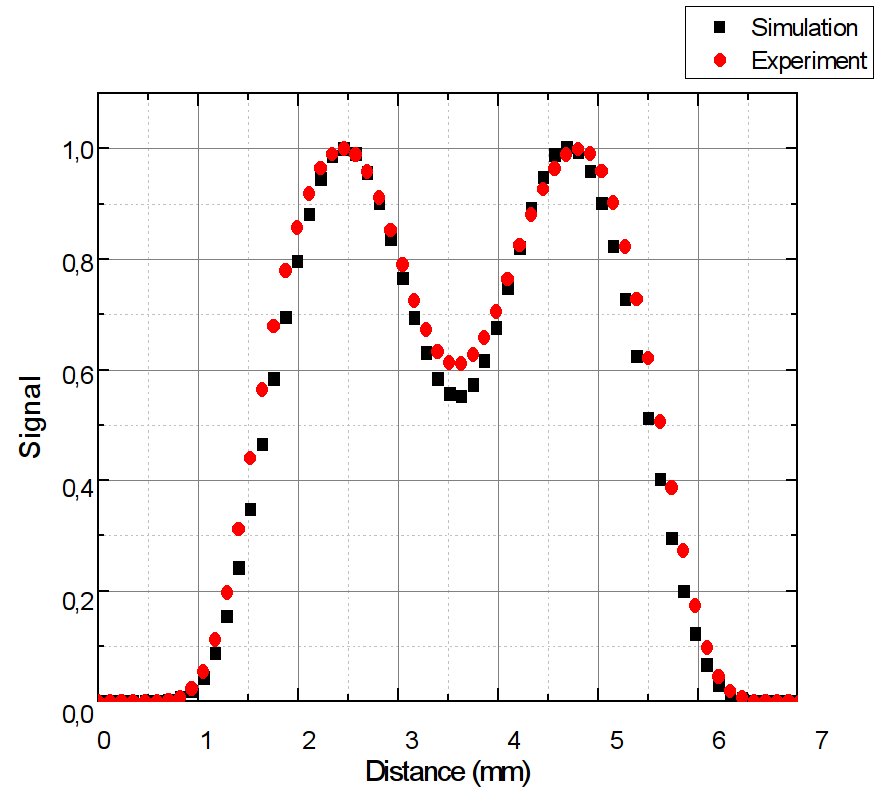}
\centering 
\caption{Comparison of simulated and experimental profiles of two $^{241}Am$ sources.}
\label{fig:11}
\end{figure}

 The comparison of the simulation and experiment using two $^{241}Am$ radioactive sources is shown in Fig.~\ref{fig:10}. The sources are clearly distinguishable, when separated from each other at a distance of 2.5~mm.  Comparison of image profiles demonstrate that the simulation is compatible with the experiment. The difference in the profile of reconstructed images (see Fig.~\ref{fig:11}) can be explained by small imperfections of the model: the intensity of the sources is assumed to be flat, the open mask elements are identical.

\par The similar comparison was made for the experiment with the capillary filled with a solution containing $^{99m}Tc$, as described above. The image profiles are shown in Fig.~\ref{fig:8}. The reconstructed image profile was fitted by a convolution of 1~mm wide rectangular distribution with a Gaussian. The spatial resolution of 0.74~mm, equal to the experimental one, was obtained in the simulation, according to the image profile. While the simulation is also compatible with the experiment, the minor difference in the tails can be associated with the different noise levels in the original shadowgrams. 

\par Finally, the simulation was used to calculate the spatial resolution and the detection efficiency for the gamma sources that are most often used in SPECT (Table~\ref{tab:4}). The point-like source was simulated. No cut on the gamma ray energy was applied during the detection efficiency calculation. 

\begin{table}[!ht]
\centering
\caption{\label{tab:4} The calculated data obtained for the gamma emitters most commonly used in SPECT.}
\smallskip
\begin{tabular}{|l|c|c|c|c|c|}
\hline
Isotope & \begin{tabular}[c]{@{}c@{}}Energy \cr[keV] \end{tabular} & \begin{tabular}[c]{@{}c@{}}Spatial \\ resolution \cr[mm] \end{tabular} & \multicolumn{2}{c|}{\begin{tabular}[c]{@{}c@{}}Registration efficiency, \% \\(with the collimator) \end{tabular}} & \begin{tabular}[c]{@{}c@{}} SNR\hspace{0.5cm}\end{tabular} \\ \cline{4-5}
&&& CdTe~1~mm~&~CdTe~2~mm&\\ \hline

$^{125}I$&	30& 	0.88&	40&	40&	96\\
\hline
$^{67}Ga$&	93.3&	0.89&	28&	36&	90\\
\hline
$^{177}Lu$&	113&	0.89&	23&	31&	88\\
\hline
$^{201}Tl$&	135&	0.89&	16&	27&	87\\
\hline
$^{99m}Tc$&	140.5&	0.89&	15&	23&	87\\
\hline
$^{117m}Sn$&	158.6&	0.90&	11&	20&	86\\
\hline
$^{123}I$&	159&	0.90&	11&	20&	85\\
\hline
$^{201}Tl$&	167&	0.90&	10&	18&	85\\
\hline
$^{111}In$&	171.3&	0.90&	10&	17&	84\\
\hline
$^{67}Ga$&	184.6&	0.91&	8&	16&	83\\
\hline
$^{177}Lu$&	210&	0.91&	7&	12&	81\\
\hline
$^{111}In$&	245.4&	0.91&	5&	10&	78\\
\hline
$^{67}Ga$&	300&	0.92&	4& 7&	74\\
\hline
$^{133}Xe$&	350&	0.92&	3&	6&	69\\
\hline
\end{tabular}
\end{table}

\begin{figure}[!ht]
\begin{minipage}[!ht]{1\linewidth}
\center{\includegraphics[scale=0.5]{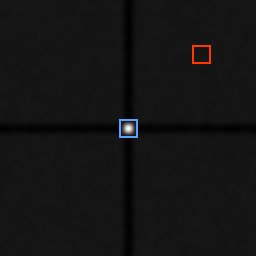}}
\end{minipage}
\caption{SNR calculation (blue square corresponds to the signal area, red square - to the background area.}
\label{fig:12}
\end{figure} 

\par The signal-to-noise ratio (SNR) was defined as follows:
\begin{equation}
\label{eq:3}
\begin{split} SNR =S/\sqrt{S^{2}+B^{2}} \end{split} ,
\end{equation}
where $S$ is the integral intensity of signal in the range of 3$\sigma$ around the maximum, and $B$ is the integral intensity of the background calculated in the similar area outside the signal image, as shown in Fig.~\ref{fig:12}. 

\par
\begin{figure}[!ht]
\begin{minipage}[!ht]{0.22\linewidth}
\center{\includegraphics[scale=0.4]{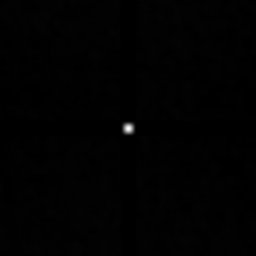}}
\end{minipage}
\hfill
\begin{minipage}[!ht]{0.22\linewidth}
\center{\includegraphics[scale=0.4]{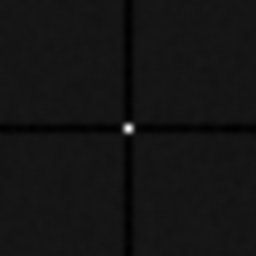}}
\end{minipage}
\hfill
\begin{minipage}[!ht]{0.22\linewidth}
\center{\includegraphics[scale=0.4]{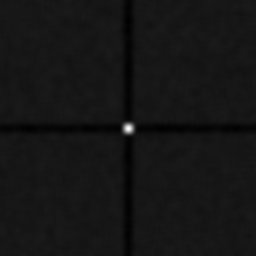}}
\end{minipage}
\hfill
\begin{minipage}[!ht]{0.22\linewidth}
\center{\includegraphics[scale=0.4]{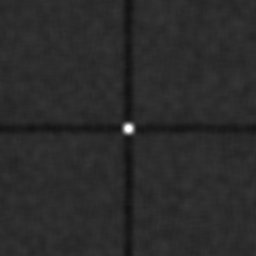}}
\end{minipage}
\vfill
\begin{minipage}[!ht]{0.22\linewidth}
\center{\includegraphics[scale=0.4]{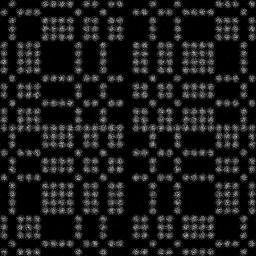}}
\end{minipage}
\hfill
\begin{minipage}[!ht]{0.22\linewidth}
\center{\includegraphics[scale=0.4]{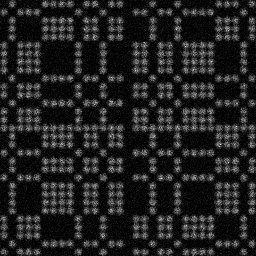}}
\end{minipage}
\hfill
\begin{minipage}[!ht]{0.22\linewidth}
\center{\includegraphics[scale=0.4]{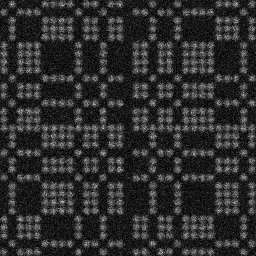}}
\end{minipage}
\hfill
\begin{minipage}[!ht]{0.22\linewidth}
\center{\includegraphics[scale=0.4]{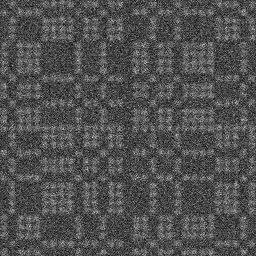}}
\end{minipage}
\caption{Simulated images of a source with the gamma ray energy of (left to right) 30, 140, 180, 350~keV. The reconstructed images are shown in the top row. The shadowgrams are shown in the bottom row.}
\label{fig:13}
\end{figure}

 It is noteworthy, that the spatial resolution slightly worsens as the photon energy increases. The higher the energy, the lower the absorption coefficient of the collimator and more photons pass thorough the opaque elements of the CA mask, thus providing a higher background counts in the detector and reducing the SNR (see Fig.\ref{fig:13}). 

\section{Conclusion}
\par A system based on the coded aperture and the hybrid pixel Timepix detector with the CdTe sensor has been used to obtain images of different gamma ray sources. The spatial resolution is shown to be of 0.8-0.9~mm at the field of view of 3~cm~$\times$~3~cm for the energy range typical for SPECT. The experimental data, supported by the simulation, demonstrate that a 1.5~mm thick tungsten coded aperture is sufficient to obtain an image of the distributed radioactive sources with the energy of gamma rays at least up to 140.5 keV without significant reconstruction artifacts. At higher energy, the image quality starts to deteriorate due to lower detection efficiency and lower absorption in the collimator. This increases the reconstruction artifacts which become evident at the energy of gamma rays about 180~keV. High spatial resolution combined with the sufficiently large field of view allows using this system for SPECT studies of small animals.
\section{Acknowledgements}
\par The reported study was funded jointly by RFBR and CITMA, project number 18-52-34005.


\begin{thebibliography}{99}
\bibitem{1}
Andringa G., Drukarch B., Bol J.G., et al., \emph{Pinhole SPECT imaging of dopamine transporters correlates with dopamine transporter immunohistochemical analysis in the MPTP mouse model of Parkinson’s disease}, Neuroimage {\bf 26} (2005) 1150.

\bibitem{2}
Zhou R., Thomas D.~H., Qiao H., et al., \emph{In vivo detection of stem cells grafted in infarcted rat myocardium}, J.~Nucl.~Med. {\bf 46} (2005) 816.

\bibitem{3}
Wild D., Behe M., Wicki A., et al., \emph{[Lys$^{\rm 40}$ (Ahx-DTPA-$^{\rm 111}$In)NH$_{\rm 2}$]exendin-4, a very promising ligand for glucagon-like peptide-1 (GLP-1) receptor targeting}, J.~Nucl.~Med. {\bf 47} (2006) 2025.

\bibitem{4}
Meikle S.~R., Kench P., Kassiou M., Banati R.~B., \emph{Small animal SPECT and its place in the matrix of molecular imaging technologies}, Phys.~Med.~Biol. {\bf 50} (2005) R45.

\bibitem{5}
Franc B.~L., Acton P.~D., Mari C., Hasegawa B.~H., \emph{Small-animal SPECT and SPECT/CT: important tools for preclinical investigation}, J. Nucl. Med. {\bf 49} (2008) 1651.

\bibitem{6}
Peterson T.~E., Furenlid L.~R., \emph{SPECT detectors: the Anger Camera and beyond}, Phys Med Biol. 2012 Sep 7; 56(17): R145–R182.

\bibitem{7}
Fontana M., Dauvergne D., Letang J.~M., Ley J.~L., Testa E. \emph{Compton camera study for high efficiency SPECT and benchmark with Anger system}, Phys. Med. Biol. {\bf 62} (2017) 8794.

\bibitem{8}
Turecek D., Jakubek J., Trojanova E., Sefc L., \emph{Compton camera based on Timepix3 technology}, JINST {\bf 13} (2018) C11022.

\bibitem{9}
Weber D., Ivanovic M., Franceschi D., et al., \emph{Pinhole SPECT: an approach to in vivo high resolution SPECT imaging in small laboratory animals}, J. Nucl. Med. {\bf 35} (1994) 342.

\bibitem{10}
Peterson T.~E., Shokouhi S., \emph{Advances in Preclinical SPECT Instrumentation}, J. Nucl. Med. {\bf 53} (2012) 841.

\bibitem{11}
Fenimore E.~E., \emph{Coded aperture imaging: predicted performance of uniformly redundant arrays},  Applied Optics {\bf 17} (1978) 3562.

\bibitem{12}
Gottesman S.~R., Fenimore E.~E., \emph{New family of binary arrays for coded aperture imaging}, Applied Optics {\bf 28} (1989) 4344.

\bibitem{13}
Llopart X., Ballabriga R., Campbell M., Tlustos L., Wong W., \emph{Timepix, a 65k Programmable Pixel Readout Chip for Arrival Time, Energy and/or Photon Counting Measurements}, Nucl. Instrum. Meth. {\bf A 581} (2007) 485.

\bibitem {14}
Procz S., Avila C., Feya J., Roque G., Schuetz M., Hamann E., \emph{X-ray and gamma imaging with Medipix and Timepix detectors in medical research}, Radiation Measurements {\bf 127} (2019) 106104.

\bibitem {15}
Accorsi R. et al., \emph{High Resolution I-125 Small Animal Imaging with a coded aperture and a hybrid pixel detector}, IEEE Trans. Nucl. Sci. {\bf 55} (2008) 481.

\bibitem{16}
Butler A., Butler P., Bell S., et al. \emph{Measurement of the energy resolution and calibration of hybrid pixel detectors with GaAs:Cr sensor and Timepix readout chip}, Phys. Part. Nucl. Lett. {\bf 12} (2015) 97.

\bibitem{17}
Greiffenberg D., Fauler A., Zwerger A., and Fiederle M., \emph{Energy resolution and transport properties of CdTe-Timepix-assemblies}, JINST {\bf 6} (2011) C01058.

\bibitem{18}
Ruat M., Ponchut C., \emph{Characterization of a Pixelated CdTe X-Ray Detector
Using the Timepix Photon-Counting Readout Chip}, IEEE Trans. Nucl. Sci. {\bf 59} (2012) 2392. 

\bibitem{19}
Zhang B., Zeng G.~L., \emph{High-Resolution Versus High-Sensitivity SPECT Imaging With Geometric Blurring Compensation for Various Parallel-Hole Collimation Geometries}, IEEE Trans. Inf. Technol. Biomed. {\bf 14} (2010) 1121.

\bibitem{20}
Van Audenhaege K., Van Holen R., Vandenberghe S., Vanhove C., \emph{Review of SPECT collimator selection, optimization, and fabrication for clinical and preclinical imaging}, Med. Phys. {\bf 42} (2015) 4796.



\bibitem{21}
Cecchin D., Poggiali D., Riccardi L., Turco P., Bui F., De Marchi S. \emph{Analytical and experimental FWHM of a gamma camera: theoretical and practical issues}, PeerJ {\bf 3} (2015) e722.

\bibitem{22}
Allison J. et al. \emph{Geant4 developments and applications}, Nucl. Instrum. Meth. {\bf A 835} (2016) 186.

\bibitem{23}
Agostinelli S. et al., \emph{GEANT4: A Simulation toolkit} Nucl. Instrum. Meth. {\bf A 506} (2003) 250.

\bibitem{24}
Apostolakis J. et al., \emph{Geant4 low energy electromagnetic models for electrons and photons} preprint CERN-OPEN-99-034.

\bibitem{25}
Cirrone G. A. P et al., \emph{Validation of the Geant4 electromagnetic photon cross-sections for elements and compounds} Nucl. Instrum. Meth. {\bf A 618} (2010) 315–322.

\end{thebibliography}
\end{document}